\documentclass[11pt,number,final]{elsarticle}
\usepackage{longtable}
\usepackage{booktabs}
\usepackage[latin2]{inputenc}
\usepackage[T1]{fontenc}
\usepackage{graphicx}
\usepackage{ae}
\usepackage{aecompl}
\usepackage{setspace}
\usepackage{amsmath}
\usepackage{amssymb}
\usepackage{amsthm}
\usepackage{natbib}
\usepackage[unicode,bookmarksopen]{hyperref}
\usepackage{todonotes}
\usepackage{parskip}
\usepackage{graphicx}
\usepackage{epstopdf}
\usepackage{float,url}
\usepackage{caption,lineno}

    \makeatletter
    \def\ps@pprintTitle{%
      \let\@oddhead\@empty
      \let\@evenhead\@empty
      \def\@oddfoot{\reset@font\hfil\thepage\hfil}
      \let\@evenfoot\@oddfoot
    }
    \makeatother

\title{Human Sexual Dimorphism of the Relative Cerebral Area Volumes in the Data of the Human Connectome Project}

\author[p]{Balázs Szalkai}
\ead{szalkai@pitgroup.org}
\author[p,u]{Vince Grolmusz\corref{cor1}}
\ead{grolmusz@pitgroup.org}
\cortext[cor1]{Corresponding author}

\address[p]{PIT Bioinformatics Group, Eötvös University, H-1117 Budapest, Hungary}
\address[u]{Uratim Ltd., H-1118 Budapest, Hungary}

\begin{document}

\begin{abstract}
The average human brain volume of the males is larger than that of the females. Several MRI voxel-based morphometry studies show that the gray matter/white matter ratio is larger in females. Here we have analyzed the recent public release of the Human Connectome Project, and by using the diffusion MRI data of 511 subjects (209 men and 302 women), we have found that the relative volumes of numerous subcortical  areas and the gray matter of most cortical areas are significantly larger in women than in men.  Additionally, we have discovered differences of the strengths of the sexual correlations between the same structures in different hemispheres.

\bigskip

\noindent Morphometry, FreeSurfer, Human Connectome Project, Sexual dimorphism, Cerebral area volumes
\end{abstract}

\maketitle

\section{Introduction}

It is known for a long time that women, on the average, have smaller brains than men (see, e.g., \cite{Witelson2006}). Therefore, if we intend to uncover sexual dimorphism of the cerebral morphometry of the sexes, the obvious measure to use is the relative size of cerebral areas, compared to the whole brain volume. With the advance of MR imaging techniques, more and more datasets have been examined and evaluated for the sex differences in the brain anatomy.

Data from forty men and forty women of ages 18-45 were analyzed in \cite{Gur1999}. It was found that men have a larger percentage of white matter and cerebro-spinal fluid than women, while the gray matter ratio was greater in women than in men. 

In a smaller study of 23 males and 23 females, \cite{Allen2002} investigated the gray- and white matter distributions in the cerebral lobes between the sexes. They have found that the volumes of the cerebral areas are larger in men, but the gray matter/ white matter ratio is higher in women.

In \cite{Chen2007} 411 subjects of ages 44-48 were examined for white matter and gray matter ratios, compared to intracranial volume. It was found that women have higher gray matter to intracranial volume than men. After a ``controlling process'', involving age, years of education, handedness and intracranial volume, some areas were found with higher relative gray matter relative volume in men and other areas with higher gray matter relative volume in women. 

In the work of \cite{Taki2011} the MRI data of 1460 healthy volunteers were examined: the regional gray matter volumes between age- and sex-groups were evaluated on 34 brain areas (\cite{Taki2011}, Figure 9 and 10) and in the whole brain (\cite{Taki2011}, Figures 3 and 4). The white matter ratio was found larger in women in all age groups while the CSF space ratio was larger in men. The gray matter ratio was found to be greater in young men than in young women; then it became larger in older women than in older men. 

In the present study we are using the dataset of one of the largest and perhaps one of the most reliable human brain study to date, the NIH-sponsored Human Connectome Project's \cite{McNab2013} Subject-500 Data Release \url{http://www.humanconnectome.org/documentation/S500}. The set contains the MRI data of 511 subjects of ages 22-35, comprising 302 women and 209 men. 

In our present study, we have compared the sizes of the brain areas between the sexes relative to the whole brain volume, without any ``controlling'' step for age, handedness or education. We have found that women have larger relative volumes for most cerebral areas.

While other studies have attempted making conclusions to brain functions from these statistically proven anatomical dimorphisms, we feel that such conclusions require more elaborate studies than just the volume comparisons (e.g., functional- and anatomical connectomics as in \cite{Szalkai2015, Szalkai2015c,Szalkai2016a}). Therefore, those conclusions related to brain functions are outside of the scope of the present work.

\section{Discussion and Results} 

The volume of different brain areas was computed as described in the ``Methods'' section. Since we are interested in the relative volume of those areas, these values were divided by the FS\_Mask\_Vol (Freesurfer's Total Brain Mask Volume). The anonymized subject-level data of the relative volumes can be downloaded as a large table from \url{http://uratim.com/relvol/tables.zip}. The first column of Table S1 contains the subject ID, the sex is given in column 2 (1 - male, 2 - female; note that the sexes were numbered this way because it was anticipated that most of the relative volumes would be larger in the case of females; so the correlations of these volumes with the quantity describing the sex would be positive).

We remark that in the third column of Table S1 one can find a few values larger than 1. The reason for this could be that the value of  FS\_InterCranial\_RelVol depends on the HCP's FS\_InterCranial\_Vol quantity, which is computed by FS talairach transform which sometimes produce unreliable results. Because of this minute unreliability we have chosen the more straightforward FS\_Mask\_Vol (Freesurfer's Total Brain Mask Volume \cite{Fischl2012}) for quantifying the ``whole'' brain volume, in the computation of the relative volumes.

Table S1 describes the subject-level relative volumes of 115 brain areas (Regions of Interests, or ``ROIs'', and larger areas as well); the brain areas are named and abbreviated as given in the documentation of the Human Connectome Project: \url{https://wiki.humanconnectome.org/display/PublicData/HCP+Data+Dictionary+Public-+500+Subject+Release}. In cortical regions the gray matter volume is considered.

Our Table 1 summarizes the most significant findings of ours: very probably, all of these correlations with the sex holds (i.e., the statistical second degree error, approximated by the Holm-Bonferroni method, is also small, i.e., it is less than 0.05). Our Table 2 lists those correlations that one-by-one have p-values less than 0.05, but their corrected p-values are larger than 0.05; that is, every single correlation in Table 2 probably holds, but it is improbable that all of them hold. 

\begin{table}
{\scriptsize
\begin{tabular}{ | l | l | l | l | l | l | }
		\hline
	{\bf Name of area} & {\bf Correlation} & {\bf p-value} & {\bf p (corr.)} & {\bf Average}  & {\bf Average}  \\ 
	 &  &  &  & {\bf (males)} & {\bf (females)} \\
	\hline
	FS\_SubCort\_GM\_RelVol & 0.3564 & 0 & 0 & 0.0364 & 0.03797 \\ \hline
	FS\_L\_Superiorparietal\_RelVol & 0.311 & 0 & 0 & 0.00734 & 0.00781 \\ \hline
	FS\_R\_Temporalpole\_RelVol & 0.2787 & 0 & 0 & 0.00094 & 0.00101 \\ \hline
	FS\_L\_Temporalpole\_RelVol & 0.2766 & 0 & 0 & 0.001 & 0.00108 \\ \hline
	FS\_Total\_GM\_RelVol & 0.2749 & 0 & 0 & 0.41307 & 0.4218 \\ \hline
	FS\_L\_Caudate\_RelVol & 0.2704 & 0 & 0 & 0.00226 & 0.0024 \\ \hline
	FS\_R\_Hippo\_RelVol & 0.2665 & 0 & 0 & 0.00266 & 0.00277 \\ \hline
	FS\_CC\_Posterior\_RelVol & 0.2599 & 0 & 0 & 0.00057 & 0.00062 \\ \hline
	FS\_R\_ThalamusProper\_RelVol & 0.2561 & 0 & 0 & 0.00442 & 0.00461 \\ \hline
	FS\_R\_Caudate\_RelVol & 0.248 & 0 & 0 & 0.00234 & 0.00247 \\ \hline
	FS\_L\_ThalamusProper\_RelVol & 0.2461 & 0 & 0 & 0.00506 & 0.00526 \\ \hline
	FS\_L\_Hippo\_RelVol & 0.2376 & 0 & 0 & 0.00262 & 0.00273 \\ \hline
	FS\_R\_Pallidum\_RelVol & 0.2258 & 0 & 0 & 0.00089 & 0.00094 \\ \hline
	FS\_CC\_Anterior\_RelVol & 0.2149 & 0 & 0.0001 & 0.00053 & 0.00057 \\ \hline
	FS\_CC\_MidPosterior\_RelVol & 0.2146 & 0 & 0.0001 & 0.00027 & 0.0003 \\ \hline
	FS\_R\_Superiorparietal\_RelVol & 0.2117 & 0 & 0.0002 & 0.00757 & 0.00789 \\ \hline
	FS\_LCort\_GM\_RelVol & 0.2024 & 0 & 0.0005 & 0.15162 & 0.15468 \\ \hline
	FS\_TotCort\_GM\_RelVol & 0.2014 & 0 & 0.0005 & 0.30691 & 0.31254 \\ \hline
	FS\_R\_VentDC\_RelVol & 0.1934 & 0 & 0.0012 & 0.00254 & 0.00263 \\ \hline
	FS\_BrainStem\_RelVol & 0.1929 & 0 & 0.0013 & 0.0131 & 0.01357 \\ \hline
	FS\_R\_Putamen\_RelVol & 0.188 & 0 & 0.0021 & 0.00336 & 0.0035 \\ \hline
	FS\_L\_VentDC\_RelVol & 0.1872 & 0 & 0.0023 & 0.00251 & 0.0026 \\ \hline
	FS\_R\_Medialorbitofrontal\_RelVol & 0.1863 & 0 & 0.0025 & 0.00303 & 0.00314 \\ \hline
	FS\_RCort\_GM\_RelVol & 0.1861 & 0 & 0.0025 & 0.15529 & 0.15786 \\ \hline
	FS\_L\_Paracentral\_RelVol & 0.1849 & 0 & 0.0028 & 0.002 & 0.0021 \\ \hline
	FS\_L\_Pallidum\_RelVol & 0.165 & 0.0002 & 0.0182 & 0.00081 & 0.00086 \\ \hline
	FS\_3rdVent\_RelVol & -0.164 & 0.0002 & 0.0196 & 0.00048 & 0.00044 \\ \hline
	FS\_L\_Superiorfrontal\_RelVol & 0.1572 & 0.0004 & 0.0355 & 0.01254 & 0.01286 \\ \hline
	FS\_L\_Putamen\_RelVol & 0.1566 & 0.0004 & 0.0369 & 0.00331 & 0.00345 \\ \hline
\end{tabular}}
\caption{The correlations of the relative volumes with the sex of the subjects. Positive correlations show that the value is greater at females, negative correlations correspond to larger relative volumes at males. The third column contains the p-values, fourth column the Holm-Bonferroni correction of the p-values \cite{Holm1979}, and the last two columns the averages for male and female subjects. This table contains data with Holm-Bonferroni corrected p-values of less, than 0.05. Table S2 at \url{http://uratim.com/relvol/tables.zip} contains the data of both Tables 1 and 2 together with more rows and standard deviation columns. The abbreviations are resolved at \url{https://wiki.humanconnectome.org/display/PublicData/HCP+Data+Dictionary+Public-+500+Subject+Release}.}
\end{table}

In Tables 1 and 2 the positive correlations show that the relative volume of the cerebral area is larger in women than in men. 

In Table 1, the highest correlation with the female sex is shown with the relative volume of the sub-cortical gray volumes. This phenomenon was described, for example, in \cite{Sowell2002}, our contribution here is the quantification of this correlation.

The left superior-parietal relative gray matter volume correlates stronger with the female sex than the right superior-parietal gray matter volume. This asymmetry is missing in the right- and left temporal pole relative volumes: both are larger in women. The relative size-advantage in the temporal lobes was described in \cite{Sowell2002}, too; we show here that there are no difference in correlations in the left- and the right hemispheres in the temporal area.

Parts of basal ganglia are also relatively significantly larger in women in both hemispheres. In the case of putamen, the correlation with sex is stronger in the right hemisphere than in the left, while in the caudate nucleus, the correlation with sex is larger in the left hemisphere than in the right one. 

The relative volume of the right pallidum is stronger correlated with the sex than the left pallidum.

The relative volume of the hippocampus is more definitely correlated with the female sex in the right hemisphere than in the left. As it was recognized by many previous studies \cite{Allen2002,Chen2007,Gur1999}, the relative volume of the total gray matter is larger in females. Now we have found that this correlation in the left hemisphere is stronger. 

In Table 1 only the relative volume of the third ventricle is larger in males than in females.

In Table 2 those results are summarized that have larger than 0.05 Holm-Bonferroni corrected p-values: that is, they could hold one-by-one with a fairly high probability (i.e., > 0.95), but it is unlikely that {\em all of them} hold true. In Table 2 the total and the right- and the left white matter ratios are larger in men than in women, as well as the relative volumes of optic chiasm, and the rostral anterior cingulate cortex, the isthmus of cingulate gyrus, the right and the left lateral ventricles.

\begin{table}
	{\scriptsize
\begin{tabular}{ | l | l | l | l | l | }
	\hline
	{\bf Name of area} & {\bf Correlation} & {\bf p-value}  & {\bf Average}  & {\bf Average}  \\ 
	&  &  &   {\bf (males)} & {\bf (females)} \\ \hline
	FS\_CC\_Central\_RelVol & 0.1525 & 0.0006 & 0.000294 & 0.000313 \\ \hline
	FS\_R\_Postcentral\_RelVol & 0.1522 & 0.0006 & 0.005333 & 0.005495 \\ \hline
	FS\_R\_Frontalpole\_RelVol & 0.1511 & 0.0006 & 0.000476 & 0.000499 \\ \hline
	FS\_L\_Postcentral\_RelVol & 0.1503 & 0.0007 & 0.005500 & 0.005656 \\ \hline
	FS\_L\_Parsopercularis\_RelVol & 0.1463 & 0.0010 & 0.002797 & 0.002914 \\ \hline
	FS\_R\_Parsorbitalis\_RelVol & 0.1367 & 0.0020 & 0.001346 & 0.001390 \\ \hline
	FS\_L\_Lateralorbitofrontal\_RelVol & 0.1331 & 0.0027 & 0.004481 & 0.004578 \\ \hline
	FS\_BrainSeg\_RelVol & 0.1299 & 0.0034 & 0.716413 & 0.721421 \\ \hline
	FS\_OpticChiasm\_RelVol & -0.1266 & 0.0043 & 0.000149 & 0.000141 \\ \hline
	FS\_CC\_MidAnterior\_RelVol & 0.1247 & 0.0049 & 0.000292 & 0.000307 \\ \hline
	FS\_L\_Posteriorcingulate\_RelVol & 0.1231 & 0.0055 & 0.001871 & 0.001929 \\ \hline
	FS\_L\_Parsorbitalis\_RelVol & 0.1228 & 0.0056 & 0.001058 & 0.001089 \\ \hline
	FS\_L\_Frontalpole\_RelVol & 0.1227 & 0.0056 & 0.000343 & 0.000357 \\ \hline
	FS\_R\_Lingual\_RelVol & 0.1199 & 0.0068 & 0.004066 & 0.004186 \\ \hline
	FS\_L\_Rostralanteriorcingulate\_RelVol & -0.1131 & 0.0107 & 0.001641 & 0.001586 \\ \hline
	FS\_L\_Parahippocampal\_RelVol & 0.1122 & 0.0114 & 0.001181 & 0.001216 \\ \hline
	FS\_L\_Pericalcarine\_RelVol & 0.1114 & 0.0119 & 0.001750 & 0.001813 \\ \hline
	FS\_R\_LatVent\_RelVol & -0.1101 & 0.0130 & 0.003885 & 0.003478 \\ \hline
	FS\_R\_Caudalanteriorcingulate\_RelVol & 0.1080 & 0.0148 & 0.001221 & 0.001280 \\ \hline
	FS\_R\_Pericalcarine\_RelVol & 0.1075 & 0.0153 & 0.001930 & 0.001996 \\ \hline
	FS\_R\_Bankssts\_RelVol & 0.1071 & 0.0156 & 0.001633 & 0.001683 \\ \hline
	FS\_R\_WM\_RelVol & -0.1069 & 0.0159 & 0.137525 & 0.135985 \\ \hline
	FS\_Tot\_WM\_RelVol & -0.1052 & 0.0177 & 0.273035 & 0.270053 \\ \hline
	FS\_L\_WM\_RelVol & -0.1024 & 0.0209 & 0.135510 & 0.134068 \\ \hline
	FS\_L\_Precentral\_RelVol & 0.1014 & 0.0222 & 0.007940 & 0.008070 \\ \hline
	FS\_L\_Lingual\_RelVol & 0.1011 & 0.0226 & 0.004012 & 0.004117 \\ \hline
	FS\_R\_Precentral\_RelVol & 0.0994 & 0.0250 & 0.008054 & 0.008179 \\ \hline
	FS\_R\_Lateralorbitofrontal\_RelVol & 0.0953 & 0.0316 & 0.004419 & 0.004487 \\ \hline
	FS\_WM\_Hypointens\_RelVol & -0.0950 & 0.0321 & 0.000546 & 0.000503 \\ \hline
	FS\_L\_LatVent\_RelVol & -0.0937 & 0.0345 & 0.004137 & 0.003736 \\ \hline
	FS\_R\_Posteriorcingulate\_RelVol & 0.0937 & 0.0345 & 0.001898 & 0.001944 \\ \hline
	FS\_R\_Rostralanteriorcingulate\_RelVol & -0.0912 & 0.0397 & 0.001267 & 0.001226 \\ \hline
	FS\_L\_Precuneus\_RelVol & 0.0896 & 0.0432 & 0.005863 & 0.005961 \\ \hline
	FS\_R\_Paracentral\_RelVol & 0.0895 & 0.0435 & 0.002356 & 0.002411 \\ \hline
	FS\_R\_Parsopercularis\_RelVol & 0.0881 & 0.0470 & 0.002414 & 0.002475 \\ \hline
	FS\_L\_Isthmuscingulate\_RelVol & -0.0877 & 0.0479 & 0.001520 & 0.001475 \\ \hline
\end{tabular}
\caption{The correlations of the relative volumes with the sex of the subjects. For these values the  Holm-Bonferroni corrected p-values are greater than 0.05. Positive correlations show that the value is greater at females, negative correlations correspond to larger relative volumes at males. The third column contains the p-values,  and the last two columns the averages for male and female subjects. Table S2 \url{http://uratim.com/relvol/tables.zip} contains the data of both Tables 1 and 2 together with more rows and standard deviation columns. The abbreviations are resolved at \url{https://wiki.humanconnectome.org/display/PublicData/HCP+Data+Dictionary+Public-+500+Subject+Release}.}}
\end{table}

\section{Methods}

For subcortical areas the HCP data release table already contained a single Volume which did not have to be post-processed. Cortical gray matter volumes were computed by multiplying the Thickness and Area data, which were available in the HCP database \cite{McNab2013}.  We then divided these volumes by FS\_Mask\_Vol (Freesurfer Total Brain Mask Volume) \cite{Fischl2012} and defined the relative volume of an ROI as this ratio. 

\subsection{Statistical analysis} We calculated Pearson's correlation coefficient \cite{Wonnacott1972} for Gender and the relative volume of each ROI. We then calculated the two-tailed p-value from this correlation coefficient $r$ and the number of subjects by using Student's t-test \cite{Wonnacott1972}. After that, we sorted the p-values from smallest to largest and used the Holm-Bonferroni method \cite{Holm1979} to correct for multiple comparisons. We have applied a threshold of 5\% for these corrected p-values.

\subsection{Data availability} The public release of the Human Connectome Project data is available at \url{http://www.humanconnectome.org/documentation/S500}. The subject-level data is available as Table S1 and the results of the statistical analysis as the Table S2 at \url{http://uratim.com/relvol/tables.zip}.
		
\section{Conclusions} We have analyzed the sexual dimorphisms of cerebral areas, applying the data of the Human Connectome Project \cite{McNab2013}. We have found that for most cortical and sub-cortical areas, the relative volumes are significantly larger in women. Additionally, we have discovered differences in the strengths of the correlations between the same structures in different hemispheres.

\section*{Acknowledgments}
Data were provided in part by the Human Connectome Project, WU-Minn Consortium (Principal Investigators: David Van Essen and Kamil Ugurbil; 1U54MH091657) funded by the 16 NIH Institutes and Centers that support the NIH Blueprint for Neuroscience Research; and by the McDonnell Center for Systems Neuroscience at Washington University.

	

\end{document}